\documentclass{elsevier-book}

\usepackage{graphicx,url}
\usepackage{array,booktabs,threeparttable,longtable}
\usepackage{multirow}
\usepackage{cite}
\usepackage{enumitem}
\usepackage{algorithm,algorithmic}
\usepackage{subfigure}
\usepackage{listings}
\usepackage{svg}
\lstdefinelanguage{JavaScript}{
  keywords={break, case, catch, continue, debugger, default, delete, do, else, false, finally, for, function, if, in, instanceof, new, null, return, switch, this, throw, true, try, typeof, var, void, while, with},
  morecomment=[l]{//},
  morecomment=[s]{/*}{*/},
  morestring=[b]',
  morestring=[b]",
  sensitive=true
}
\usepackage{amsmath,amssymb}

\usepackage{amsthm}

\theoremstyle{remark}

\begin{document}
\chapter{Quantum Federated Learning: Architectural Elements and Future Directions}

\chapterauthors{Siva Sai, Abhishek Sawaika, Prabhjot Singh and Rajkumar Buyya}{%
Quantum Cloud and Distributed Systems (qCLOUDS) Lab, \\
School of Computing and Information Systems, \\
The University of Melbourne, Australia\\[0.3em]
}

\begin{abstract}
Federated learning (FL) focuses on collaborative model training without the need to move the private data silos to a central server. Despite its several benefits, the classical FL is plagued with several limitations, such as high computational power required for model training(which is critical for low-resource clients),  privacy risks, large update traffic, and non-IID heterogeneity. This chapter surveys a hybrid paradigm - Quantum Federated Learning (QFL), which introduces quantum computation, that addresses multiple challenges of classical FL and offers rapid
computing capability while keeping the classical orchestration intact. Firstly, we motivate QFL with a concrete presentation on pain points of classical FL, followed by a discussion on a general architecture of QFL frameworks specifying the roles of client and server, communication primitives and the quantum model placement. We classify the existing QFL systems based on four criteria - quantum architecture (pure QFL, hybrid QFL), data processing method (quantum data encoding, quantum feature mapping, and quantum feature selection
\& dimensionality reduction), network topology (centralized, hierarchial, decentralized), and quantum security mechanisms (quantum key distribution, quantum homomorphic encryption, quantum differential privacy, blind quantum computing). We then describe applications of QFL in healthcare, vehicular networks, wireless networks, and network security, clearly highlighting where QFL improves communication efficiency, security, and performance compared to classical FL. We close with multiple challenges and future works in QFL, including extension of QFL beyond classification tasks, adversarial attacks, realistic hardware deployment, quantum communication protocols deployment, aggregation of different quantum models, and quantum split learning as an alternative to QFL. This chapter helps in understanding and evaluating QFL systems in different theoretical and application scenarios. 
\end{abstract}


\section{Introduction}
Federated Learning \cite{zhang2021survey} trains machine learning models across several data silos without the need to move the raw data to a central server. In a typical FL iteration, the institutional nodes or clients download a shared global model, train the model using their private datasets, and share only the updates to a coordinating server. The server aggregates the updates from the clients and broadcasts them back for the next round. The federated learning setup allows learning from data that would be impossible due to privacy regulations, ownership constraints and bandwidth. It also lowers privacy risks and avoids the single point of failure problem. Despite the several benefits, FL meets a set of scientific and practical requirements at scale, which expose limitations in classical frameworks and consequently open a path towards quantum enhancements. 
A few of the relevant challenges in the FL setups are as follows:
\begin{enumerate}[]
    \item High compute requirement: Classical FL faces several practical constraints in low-resource clients scenario, inlcuding high computational and memory demands of local model training, and the consequent latency and energy burdens.     
    \item Privacy and security issues: The core reason behind the privacy preservation in FL settings is that the clients only share the gradients without sharing the raw data. But that doesn't eliminate leakage completely. Membership inference, model inversion and poisoning by the adversaries still limit the privacy, thus raising the need for stronger protocols to resist adversaries and bound leakage.
    \item High-dimensional and multi-modal data: Several application domains of FL, including healthcare, manufacturing and finance, produce data with several features and modalities like text, image, and temporal data. Hence, there is a clear need for models with the capability of powerful representation learning.
    \item Communication efficiency: In the FL setting, frequent exchange of large model updates creates heavy downlink and uplink traffic. Although the techniques like sparsification and compression help, the tradeoff between bandwidth and accuracy remains a strong factor. 
    \item Heterogeneity and statistical robustness: Real-world data in an FL setting is almost always non-IID. The client's data may have different label distributions, sample sizes, and feature spaces. This heterogeneity leads to issues like instability during aggregation, model bias, and slow convergence. 
\end{enumerate}
The integration of Quantum Computing (QC) into FL addresses several of these concerns effectively. 
\subsection{Motivation for Quantum Federated Learning (QFL)}
The integration of quantum computing and federated learning is based on the fact that each of these technologies supports the other. The integration of quantum computing into federated learning environments offers notable benefits, but that is not to minimize the use of the federated learning paradigm in quantum computing setups.  

QC provides exceptional computational capabilities, superseding the classical systems. QC enables unique algorithms capable of solving complex problems which are intractable to solve with classical systems. Non-convex loss surfaces are typical in deep learning, which leads classical FL systems to struggle with complex optimization problems, especially when dealing with large-scale data. Quantum optimization algorithms enabled by QC explore large solution spaces very efficiently by leveraging quantum parallelism. A few notable optimization algorithms include Quantum Approximate Optimization (QAOA) and Variational Quantum Eigensolver (VQE). These quantum algorithms lead to improved accuracy and convergence as they can potentially find better minima in high-dimensional loss spaces. Classical FL fails to handle quantum data or systems based on quantum mechanics, like quantum sensor networks and quantum chemistry. QC is a natural option to train QML models in these cases to represent and manipulate quantum states. Thus, by its integration into FL systems,  QC enlarges the application space of FL into quantum-domain tasks. QC also assists FL in achieving reduced communication bottlenecks through quantum compression and encoding. A main limitation of classical FL is communication overhead due to recurrent transmission of large model updates. QC-enabled techniques like entanglement-assisted compression and quantum teleportation can significantly reduce the quantum information that needs to be shared across nodes\cite{li2023entanglement}. 

Quantum Machine Learning (QML) models like Quantum Neural Networks (QNNs) and Variational Quantum Circuits (VQCs) can define highly non-linear entangled representations and feature maps with relatively few parameters. This would be very helpful for the clients with a limited computation budget in the QFL setting. Quantum generative models are very efficient in representing complex distributions. In the FL setting, the quantum generative models could be exploited for calibration or data augmentation of skewed local datasets while keeping raw data private. QC can help FL frameworks strengthen their privacy by incorporating quantum protocols like blind quantum computing, which enables computation on quantum-encrypted data. 

FL becomes particularly important in privacy-sensitive quantum domains such as wireless networks and IoT. It also enables collaborative and scalable learning across distributed devices. Centralized quantum machine learning faces scalability challenges due to the need for transmission of large volumes of intermediate states and quantum data to a central processor. Current quantum communication infrastructure makes it almost impossible \cite{10856200}. By incorporating FL, only the quantum model updates would be shared among the participants, thus making building intelligent quantum systems efficient. QFL would power a broad range of real-world quantum-driven applications while simultaneously improving data privacy. 

\subsection{Chapter Organization}
The rest of the chapter is organized as follows. Having covered the motivation of integration of quantum computing and federated learning in the introduction, the next section talks about the general QFL architecture. After that, we present a taxonomy of QFL systems - classified based on four important criteria. Then, we analyze the applications of QFL in a few highly relevant domains. After that, we explore current challenges and future directions in QFL. Finally, we conclude the chapter with key insights. 

\section{Background}

\subsection{Power of quantum computing}

The fundamental unit of classical information is the bit, which can exist in only two states: 0 or 1. Conversely, the qubit, the basic unit of quantum information, can exist in a superposition of orthogonal states \(|0\rangle\) and \(|1\rangle\). A superposition is a state in between \(|0\rangle\) and \(|1\rangle\), represented as \(\alpha|0\rangle + \beta|1\rangle\), where \(\alpha\) and \(\beta\) are complex numbers and are called amplitudes, making this different from a probabilistic state as it has real numbers as amplitudes. One other property of a qubit is that upon measurement, it collapses to one of its orthogonal states. The probability that the state collapses to either orthogonal state is determined by the norm square of the amplitude of the respective state.
In a quantum circuit, the Hadamard gate facilitates the introduction of superposition. Superposition enables us to perform operations on multiple states simultaneously; on the other hand, when we measure, the state collapses and we get a single value. This makes it very hard to create quantum algorithms, so interference effects are used to ensure the correct answer is reached.
Another crucial aspect introduced by qubits is entanglement, a phenomenon where two systems, despite their physical separation, exhibit correlated behaviours that defy classical randomness. In a quantum circuit, the CNOT gate enables the introduction of superposition.

The power of superposition and entanglement allows the development of algorithms that deliver exponential speedups in specific scenarios.
One notable example of this is the Deutsch-Jozsa Algorithm.
The primary objective of this algorithm is to ascertain whether a function exhibits constant behavior or exhibits balance. Consider a hypothetical black box (or an Oracle) that accepts input as a n-bit string and outputs either 0 or 1. Classically, it would require at most \(N/2 + 1\) queries to determine if the function is constant or balanced, where \(N = 2^n\).
However, one can achieve this determination in a single query using the power of quantum computing. Initially, an equal superposition of all possible \(N\) states is constructed by applying Hadamard gates to each of n qubits. Subsequently, this information  is passed through a phase oracle (an oracle analogous to the classical oracle, ensuring its reversibility, as quantum gates must be reversible). Finally, Hadamard gates are applied to each qubit once more, resulting in the final state.
Equation 1 illustrates the process:
\begin{equation}
    \sum_{z \in\{0,1\}^{n}}\left(\frac{1}{2^{n}} \sum_{x \in\{0,1\}^{n}}(-1)^{f(x)+x \cdot z}\right)|z\rangle
\end{equation}
Upon performing measurement, the probability of obtaining all zeros is 

\(\left(\frac{1}{2^{n}} \sum_{x \in\{0,1\}^{n}}(-1)^{f(x)}\right)^2\). If the function is constant, \((-1)^{f(x)}\) remains constant, resulting in a probability of 1. Conversely, if the function exhibits balance, \((-1)^{f(x)}\) alternates between +1 and -1, yielding a probability of 0.
Therefore, by measuring the n qubits, one can determine whether the function is constant or balanced. If all zeros are obtained, the function is constant; otherwise, it is balanced.
Apart from the Deutsch-Jozsa algorithm, other algorithms such as Bernstein-Vazirani, Simon, and Grover algorithms provide significant speedup compared to their classical counterparts, demonstrating the immense power of quantum computing.

\subsection{Common tools used in quantum computing}
\begin{itemize}
    \item \textbf{Qiskit} – IBM’s open-source SDK for building, simulating, and running quantum circuits on real IBM Quantum hardware.
    \item \textbf{PennyLane} – A Python library for differentiable quantum programming and hybrid quantum-classical machine learning.
    \item \textbf{Cirq} – Google’s framework for designing, simulating, and executing quantum circuits on Google Quantum processors.
    \item \textbf{Braket SDK} – Amazon’s Python interface to build, test, and run quantum algorithms on cloud-hosted hardware backends.
    \item \textbf{QuTiP} – Quantum Toolbox in Python for simulating open quantum systems and dynamics using density matrices.
    \item \textbf{t\textbar ket$\rangle$} – Cambridge Quantum’s compiler optimizing quantum circuits for different hardware architectures.
    \item \textbf{Ocean SDK} – D-Wave’s developer suite for formulating and solving problems on quantum annealers.
    \item \textbf{Q\# / QDK} – Microsoft’s domain-specific language and development kit for quantum algorithm design and simulation.
    \item \textbf{Strawberry Fields} – Xanadu’s platform focused on photonic quantum computing and continuous-variable circuits.
    \item \textbf{ProjectQ} – A Python framework for high-performance quantum simulation and flexible backend compilation.
\end{itemize}

\subsection{Quantum Hardware}
Quantum hardware is diverse, with each architecture offering unique advantages. Superconducting transmons, used by IBM and Google, provide fast gates and well-developed tools, though they have shorter coherence times \cite{kjaergaard2020superconducting}. Trapped ions, from IonQ and Quantinuum, are known for their excellent coherence and comprehensive connectivity, albeit with slower gate speeds \cite{bruzewicz2019trapped}. Neutral-atom Rydberg systems, developed by QuEra and Atom, allow for rapid qubit scaling with adaptable layouts, but they are still refining their error rates \cite{morgado2021quantum}. Photonic platforms, like those from Xanadu, are notable for their room-temperature operation and integration with optics, though achieving universal high-fidelity gates is still a work in progress \cite{flamini2018photonic}. Spin qubits in silicon or diamond offer the potential for CMOS-style scalability but are still in the early stages for large-scale algorithms \cite{zwanenburg2013silicon}. Lastly, topological approaches focus on intrinsic error protection, with practical devices still being developed \cite{alicea2012new}.

\section{Architecture}

In this section, we present the working concept of QFL, which is illustrated in Figure \ref{fig:centralized-qfl}. The QFL paradigm enables collaborative training of a shared model by multiple quantum-enabled devices. This decentralized training approach enhances data privacy and leverages the computational advantages of quantum-enabled devices. In vanilla QFL, a central server coordinates the overall QFL process. The coordination includes management of aggregation, communication, and synchronization between quantum devices. The QFL (centralized QFL in particular) follows a multi-step and cyclical process with the following key stages:

\begin{figure*}[!t]
    \centering
    \includegraphics[width=0.8\textwidth]{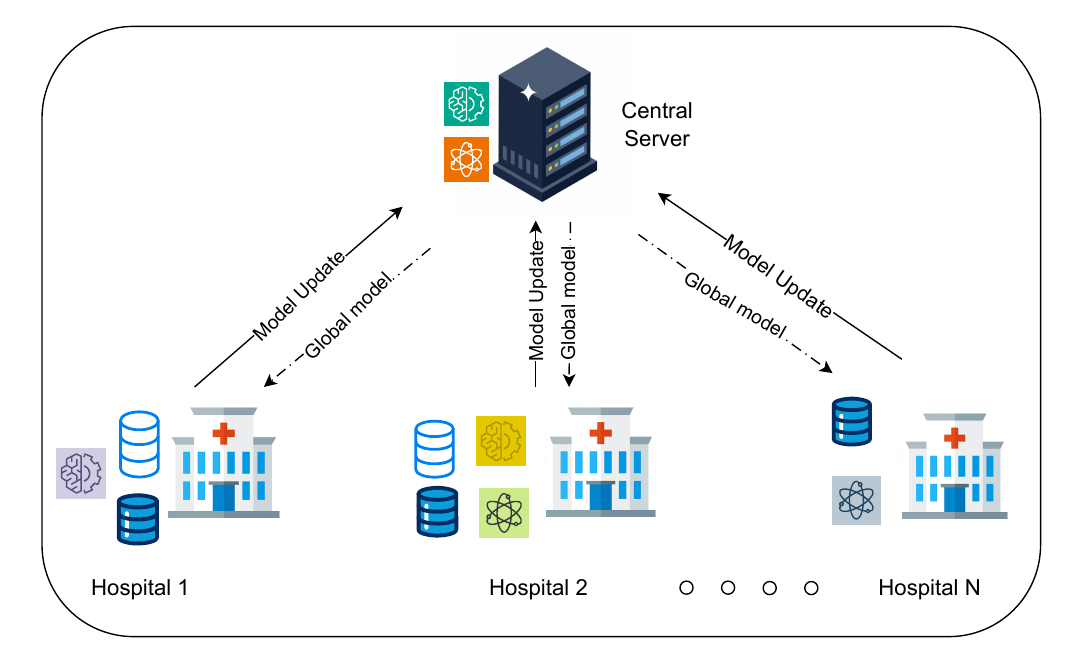}
    \caption{Centralized QFL architecture.}
    \label{fig:centralized-qfl}
\end{figure*}

\begin{enumerate}
\item Data Encoding: Each client encodes its private classical data into quantum states. A quantum state encoder is used to map classical features into quantum Hilbert space representations. State preparation is the starting step in the encoding process, where each qubit is initialized in its ground state. There are several encoding techniques that can be employed depending on the desired learning tasks and the nature of the data. Amplitude encoding, basis encoding, and angle encoding are popular choices. The data encoding steps make sure that the data is transformed into a form that is compatible with parametrized quantum circuits.
\item Local Training: After the data is encoded, it is passed through a parametrized quantum circuit at each client. Generally, this circuit is finalized by the server and passed down to all the clients. A Parametrized Quantum Circuit (PQC) consists of a set of quantum gates, the behaviour of which is controlled by trainable parameters. The quantum gates explore complex quantum correlations in the data by manipulating and entangling the qubits. Post the application of PQC, the qubits are measured, and the expectation values or measurement probabilities (resulting classical outputs) are used to compute loss functions. The client updates its local parameters using gradient-based optimization techniques like Stochastic Gradient Descent (SGD). Quantum version of SGD called Quantum Natural Gradient Descent (QGND) can also be used based on the type of circuit at the client. 

\item Aggregation and model sharing: After the local training, each client shares its updated PQC parameters or gradients to the central server. The server generates an updated global model by applying techniques like FedAvg. The server tries to ensure that the aggregation is privacy-preserving and safe. After the aggregation, the server sends back the model to every client, which is then used by the clients to update their local models. Thus, an iteration of QFL training is completed, and the same process is continued for several rounds until the models achieve an acceptable performance or converge. 

\end{enumerate} 

The QFL training process looks similar to that of classical FL; however, QFL has multiple special features, including quantum-specific noise management, circuit depth restructions, and quantum encoding. All these features make QFL an appropriate choice for privacy-sensitive use cases in quantum-enabled environments.
Schuld et al. \cite{schuld2018supervised} classify the intersection of machine learning and quantum computing into four categories - quantum for classical (processing classical data on quantum computers), quantum for quantum( processing quantum data on quantum computers), classical for quantum (processing quantum data on classical computers), and classical for classical (processing classical data on classical computers, but with quantum-inspired algorithms). While the QFL paradigm applies to all the categories, it is more relevant to quantum for classical and quantum for quantum. 
QFL includes the possibility of a hybrid nature of clients and servers(with the capabilities to run either or both of the classical and quantum neural network models), thus aligning with the current real-world scenarios where complete training on quantum devices only is infeasible.

\subsection{An approach for realization of QFL}
To establish a quantum federated learning (QFL) system, one must first implement a hybrid federated-learning pipeline. In this configuration, each client trains a local variational quantum model or a hybrid quantum–classical model and transmits model updates to a central aggregator, such as Federated Averaging (FedAvg) or Federated Proximal (FedProx), while complying with privacy and communication constraints. Tools like PennyLane offer differentiable programming with seamless integration into PyTorch, TensorFlow, and JAX, supporting gradient methods such as parameter-shift and adjoint differentiation. Qiskit provides robust transpilation, advanced noise modeling, and direct access to IBM Quantum hardware through Aer and Runtime. The pure or hybrid QML models required in QFL can be built using tools like Pennylane and Qiskit. PennyLane can also connect to various providers via plugins, including AWS Braket, IonQ, and IBM.

A staged simulation strategy should be implemented: one should begin with small-qubit analytic or statevector simulations, progress to shot-based noisy simulations (such as Qiskit Aer noise models from “Fake” backends like FakeManila), and finally conduct targeted hardware runs. In practice, large circuits with more than approximately 100 qubits are typically impractical for end-to-end QFL due to circuit depth, device noise, and queueing constraints. For orchestration, frameworks like Flower, along with FedML or TensorFlow Federated, enable the creation of custom clients that call quantum backends and return gradients or expectation values. Throughout the process, it is crucial to plan for device heterogeneity, error-mitigation techniques, secure aggregation, and bandwidth limitations.

\section{Taxonomy}
\begin{figure*}[!t]
    \centering
    \includegraphics[width=1\textwidth]{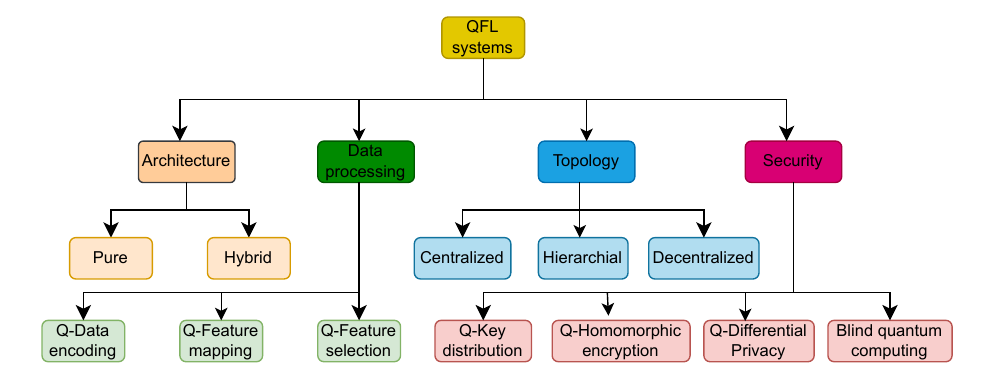}
    \caption{Classification framework for QFL systems based on architecture, data processing, topology, and security mechanism dimensions.}
    \label{fig:qfl-taxonomy}
\end{figure*}

As shown in Figure \ref{fig:qfl-taxonomy}, we classify the existing QFL systems based on four aspects -  quantum architecture \cite{comst}, data processing method, network topology, and involved security mechanism. 

\subsection{Quantum architecture}
Based on the type of machine learning model used in the federated setup, we divide the QFL models into two categories - those that incorporate pure quantum machine learning models (Pure QFL)  and those that incorporate hybrid quantum models (Hybrid QFL), i.e., combining classical neural network layers with quantum layers. 

Pure QFL enables training of a global quantum model, maintaining the data privacy of sensitive local classical or quantum data. A common configuration in Pure QFL setup is training of PQCs with the VQE optimization technique \cite{huang2022quantum}. PQCs, with modifiable classical parameters, enable quantum devices to learn and incorporate intricate patterns in complex systems. When the clients securely upload measurement results or quantum model updates, the server uses a suitable quantum aggregation technique to create a global quantum model, which is later sent to the clients. Pure QFL models make use of quantum techniques like superposition, inference, and quantum entanglement to learn from data more efficiently. As quantum features limit inferences of device data, pure QFL models tend to be more secure against unauthorized data breaches, FL system faults, and adversarial attacks. Recent research developed Pure QFL models for multiple application areas. 
Park et al. \cite{park2025entanglement} introduced entangable slimmable quantum neural networks (esQNNs) to adapt QFL to work well under changing channel conditions in IoT environments. Based on esQNNs, the authors propose an entangable slimmable QFL (esQFL) framework wherein the superposition-coded parameters of esQNNs are shared.
Huong et al. \cite{huang2022quantum} proposed a communication-efficient Variational Quantum Algorithm(VQA) and showcased the superiority of the proposed model on near-term processors. Abou et al. \cite{abou2024privacy} proposed a QFL-enabled intrusion detection system to detect network intrusions in consumer electronics networks.  

Hybrid QFL models combine quantum layers with classical neural network layers. Generally, these models consist of initial convolutional or fully connected layers to lower the computational overhead by effectively handling large-dimensional data. These initial layers take care of data processing and feature extraction. After that, the extracted features are encoded into quantum states and passed on to PQCs. The quantum circuits of further quantum layers are parametrized by classical parameters, which are trainable. Classical optimization algorithms are used to enable quantum circuits to learn optimal parameters in an iterative fashion. After the training of the hybrid QML models at the clients, they are passed on to the server for aggregation. Ren et al. \cite{ren2023qfdsa} proposed a hybrid QFL system for smart cyber-physical dynamic security assessment. Hisamori et al. \cite{hisamori2024hybrid} proposed a hybrid QFL framework, which consists of a two-qubit quantum circuit embedded with a convolutional neural network that further uses variational quantum circuits. 

\subsection{Data processing method}
When classical data is fed for training in QFL systems, it needs to be transformed in such a way that QML models can process it. Based on the data processing methods, we classify the QFL systems into three categories: quantum data encoding-based, quantum feature mapping-based, and quantum feature selection and dimensionality reduction-based \cite{tnnls}.

QFL systems employing quantum data encoding and processing methods investigate different encoding schemes with the aim of efficiently transforming the classical data into quantum states. The broader goal of these systems is the minimization of the resources required for processing and encoding the varied data(discrete, continuous, and categorical) in quantum states. Recent research developed multiple QFL systems employing quantum encoding methods. Yang et al. \cite{yang2022bert} proposed a vertical quantum federated learning architecture based on a quantum-BERT model for text classifications. The authors made use of variational quantum circuits for quantum encoding of the textual data. 
In another study, Yang et al. \cite{yang2021decentralizing} proposed a privacy-preserving decentralized feature extraction method for speech recognition. The authors upstreamed the input speech to a quantum computing server for extracting Mel-spectrogram features in a decentralized setup. They used VQC to encode the convolutional features. Huang et al. \cite{huang2022quantum} also utilize a quantum encoding method to create quantum data. 

The QFL systems based on the quantum feature mapping technique exploit quantum properties to create nonlinear and high-dimensional representations of classical data. Quantum versions of the classical feature transformation methods, like kernel methods, are developed by QFL systems in this category. Apart from developing resource-efficient quantum mapping methods, the systems also aim at building optimal quantum feature maps that balance the robustness with the expressiveness of quantum feature mapping. Yun et al. \cite{yun2022quantum} developed a quantum split leural network framework which also employs cross-channel pooling to efficiently leverage the unique quantum state tomography built by QCNN. The proposed framework achieved better performance than a traditional QFL system in terms of communication cost,  faster convergence, and privacy. 

The QFL systems employing quantum feature selection aim at identifying the most informative and relevant features by strategic selection of a subset of important features, while the quantum dimensionality reduction techniques focus on reducing the dimensionality of a dataset while preserving its important characteristics in a quantum computing environment. An important way of quantum feature selection is by analyzing the weights of the trained QNN to find the parameters that contribute better to the network's predictions. Li et al. \cite{li2022implementing} proposed a quantum variational feature selection, which is based on graph theory and the quantum approximate optimization algorithm. He et al. \cite{he2020quantum} proposed a quantum locally near embedding for nonlinear dimensionality reduction, which can be employed in QFL systems for dimensionality reduction.

\subsection{Network topology}
Network topology refers to the way the different quantum systems are connected in the QFL setup. Based on network topology, we divide the QFL systems into three categories: centralized QFL, hierarchical QFL, and decentralized QFL. 

In this most popular topology of centralized QFL systems, the central server organizes the model aggregation and coordinates the communications across the client devices based on a hub-and-spoke network model topology. Each client will have either pure quantum models or hybrid quantum models, which they train and send the updates to the server, and receive the updated global model in every iteration. Reduced system complexity, better control over training and synchronizing convergence, and simplified communication protocols are the primary benefits of the centralized QFL setup. The server may employ sophisticated quantum techniques like blind quantum computing and quantum key distribution to guard against hostile and eavesdropping interference nd to guarantee safe communication. A key disadvantage of the centralized QFL topology is that the system is susceptible to the single point of failure problem. Zhang et al. \cite{zhang2022federated} proposed a centralized QFL system based on quantum secure aggregation, which guarantees that all attempts to eavesdrop local model parameters would be immediately detected and stopped. Liu et al. \cite{zhang2022federated} proposed a practical QFL system that enables secure aggregation with information-theoretic security and utilizes distributed quantum secret keys in order to protect local model updates. The authors validated their framework experimentally based on a 4-client quantum neural network.

Hierarchical QFL systems follow a multi-tiered networking design to lower latency, improve scalability, and control quantum resource allocation over large-scale distributed systems. Under this design, the upper layer consists of the central quantum cloud server, while quantum edge servers and quantum clients constitute the lower layer. Through this tiered communication hierarchy, system scalability and training efficiency are enhanced to a great degree. The quantum clients train a local pure or hybrid QML model and send its local updates to quantum edge servers. These intermediate edge servers perform intra-cluster aggregation and then transfer the intermediate global model to the central quantum cloud server, which receives many such intermediate global models and aggregates them to get a final global model. Naorottama et al. \cite{hierarchial} proposed a two-tier hierarchical QFL framework for power allocation optimization in wireless networks. Gurung et al. \cite{gurung2025sat} also proposed a hierarchical QFL framework for secure low orbital satellite networks. 

In the decentralized QFL setups, several quantum clients collaboratively train a global model based on peer-to-peer networking, thus eliminating the need for a central server. Both the pure and hybrid QFL models can be trained with this setup. After the quantum models are locally trained on the clients, the clients directly exchange the quantum measurement results or model parameters with their neighbouring devices over safe communication links based on quantum key distribution and entanglement-assisted quantum networks. Decentralized QFL systems enhance resistance against adversarial attacks, scalability, and fault tolerance, and also remove the single point of failure problem. The quantum systems make use of distributed protocols like blockchained-inspired synchronizing techniques, quantum gossip protocols, and quantum consensus algorithms to achieve collective model convergence. 
Gurung et al. \cite{gurung2023decentralized} designed and implemented a trustworthy and decentralized QFL framework for Metaverse services. The proposed systems create transparent and secure systems that are robust to fraud and cyberattacks by leveraging blockchain technology. In another study, Gurung et al. \cite{gurung2025chained} proposed a chained continuous QFL framework which avoids the need for a server and aggregation. In the proposed model, the clients participate in a sequential continuous QFL process. They collaboratively exchange the models among themselves. 

\subsection{Quantum Security mechanism}
QFL frameworks often employ one or more quantum security mechanisms to defend against several possible attacks in the federated learning setup. Based on the employed quantum security mechanism, we divide the QFL systems into four categories: quantum key distribution-based systems, quantum homomorphic encryption-based systems, quantum differential privacy-based systems, and blind quantum computing-based systems. 

Quantum Key Distribution (QKD) protocols allow two distributed participants to generate shared cryptographic keys, which ensures secrecy is securely granted. These protocols are inspired by the laws of quantum physics. The quantum principles like quantum superposition, entanglement, and the Heisenberg uncertainty principle are exploited to grant security to QKD protocols. QKD protocols ensure detectable disruptions to the quantum systems in the case of any attempt to eavesdrop. They also rely on the fact that the quantum-generated keys are immune to both quantum and classical attacks\cite{ravikumar2023quantum}. BB84, BB92, and E91 are a few popular QKD protocols\cite{qkd}. Based on the channel noise or key refresh rate in QKD protocols, QFL systems can dynamically change communication intervals and model update rates. Thus, the use of QKD protocols in QFL systems enhances both the system performance and the security aspects. Liu et al. \cite{liu2025practical} developed a practical QFL framework that ensures secure aggregation with information-theoretic security and protects local model updates by exploiting QKD protocols. The authors validated their framework on a 4-client framework with a scalable structure. Gurung et al. \cite{gurung2025sat} proposed an access-aware and hierarchical QFL framework that partitions satellites into primary and secondary systems and schedules simultaneous, asynchronous, and sequential edge training. The authors used QKD protocols to grant quantum-resilient integrity and confidentiality, enabling authenticated encryption for model exchange. The authors also make use of quantum teleportation for quantum state transfer. 

\begin{figure}[ht]
    \centering
    \includegraphics[width=1\textwidth]{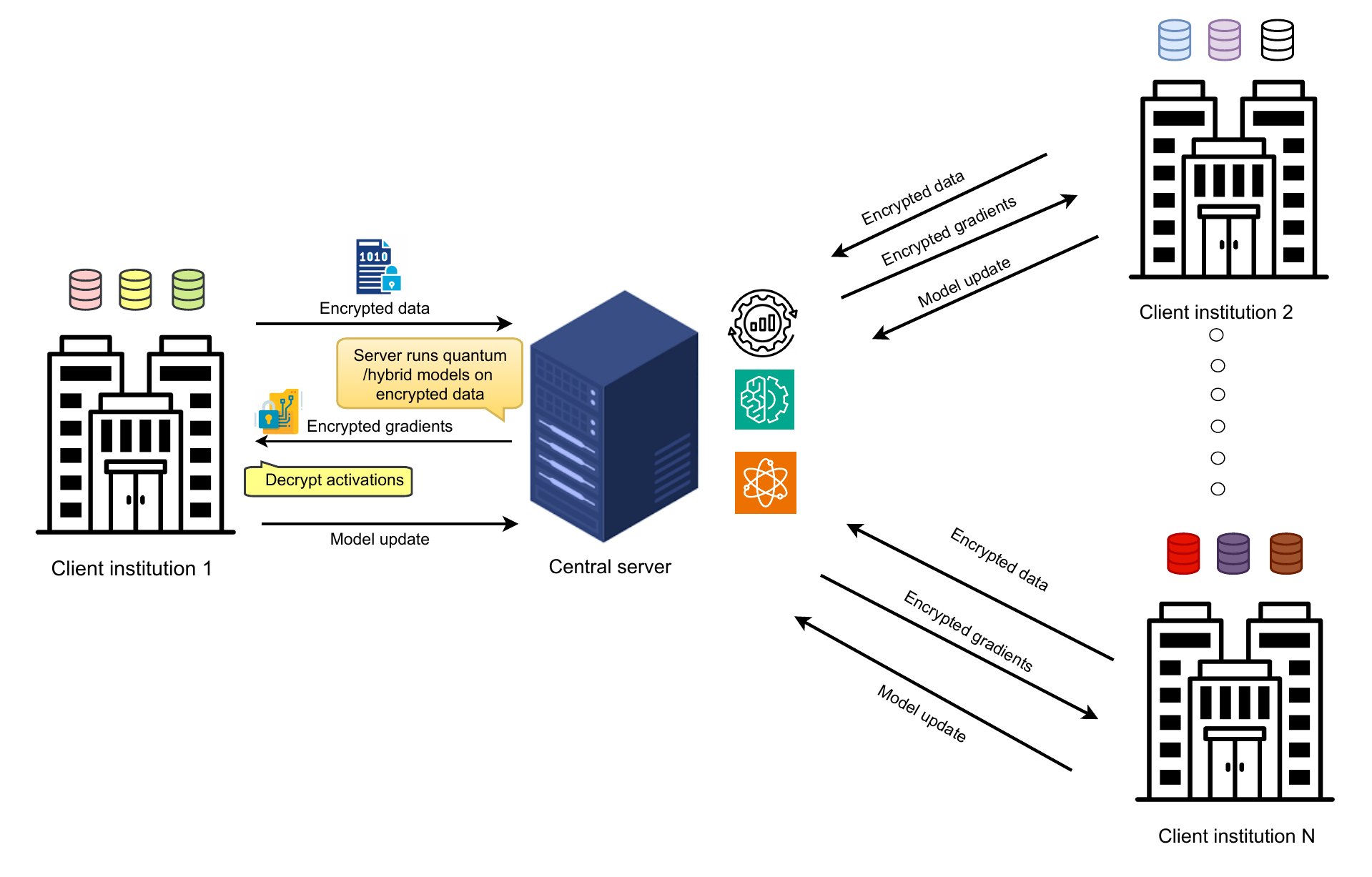}
    \caption{Quantum Homomorphic Encryption security mechanism in a QFL framework}
    \label{fig:qfl-qhe}
\end{figure}

Quantum Homomorphic Encryption (QHE) provides a quantum analog to classical homomorphic encryption, which preserves confidentiality and privacy by allowing the calculations to be performed on encrypted content without the need for decryption. Thus, QHE allows computation on encrypted quantum states and thus even guards against quantum-powered attackers. A representative diagram of a QHE-enabled QFL system is shown in Figure \ref{fig:qfl-qhe}. In QHE-enabled QFL systems, the clients securely provide encrypted updates after training on classical or quantum data to the central server, which trains a global QML model. All the computations at the server happen on encrypted quantum states. 
Chu et al. \cite{chu2023cryptoqfl} proposed CryptoQFL system that enables distributed QML training on encrypted data. CryptoQFL exploits the QHE mechanism to encrypt the local model updates before sending them to the central server. The authors also present a quantum aggregation circuit that significantly reduces the latency. Li et al. \cite{li2025quantum} also employ QHE for securing their QFL framework. 
A limitation of QHE in QFL systems is its deployability due to the existing quantum computing constraints, like controlling quantum noise and maintaining coherence. 

\begin{longtable}[!t]{|p{1cm}|p{10cm}|p{1.9cm}|p{1.85cm}|}
\caption{Summary of the covered applications of QFL}
\label{appns} \\
\hline
\textbf{Work} & \textbf{Contribution} & \textbf{Quantum Architecture} & \textbf{Topology} \\
\hline
\endfirsthead

\caption{Summary of the covered applications of QFL (continued)} \\
\hline
\textbf{Reference} & \textbf{Contribution} & \textbf{Quantum Architecture} & \textbf{Topology} \\
\hline
\endhead

\hline
\multicolumn{4}{|r|}{{Continued on next page}} \\
\hline
\endfoot

\hline
\endlastfoot

\cite{qu2025daqfl} & 
Dynamic aggregation quantum federated learning algorithm for intelligent diagnosis in Internet of Medical Things
& Pure
& Centralized \\
\hline

\cite{balasubramani2025novel} &
Framework for quantum-enhanced federated learning with edge computing for advanced pain assessment
& Hybrid
& Centralized \\
\hline

\cite{bhatia2023federated} &
Federated quanvolutional neural network in a healthcare scenario
& Hybrid
& Centralized \\
\hline

\cite{bhatia2024federated} &
Federated hierarchical tensor networks for healthcare
& Hybrid
& Centralized \\
\hline

\cite{hazarika2025quantum} &
Quantum-enhanced federated learning for metaverse-empowered vehicular networks
& Pure
& Decentralized \\
\hline

\cite{yamany2023oqfl} &
Quantum-based federated learning framework for defending against adversarial attacks in intelligent transportation systems
& Hybrid
& Centralized \\
\hline

\cite{qu2024qfsm} &
Quantum federated learning algorithm for speech emotion recognition in 5G IoV
& Hybrid
& Centralized \\
\hline

\cite{quy2024federated} &
Quantum federated learning for space-air-ground integrated networks
& Pure
& Centralized \\
\hline

\cite{chaudhary2024quantum} &
Quantum federated reinforcement-learning-based joint mode selection and resource allocation for STAR-RIS-aided VRCS
& Pure
& Centralized \\
\hline

\cite{subramanian2024hybrid} &
Hybrid quantum enhanced federated learning for cyber attack detection
& Hybrid
& Hierarchical \\
\hline

\cite{elmaouaki2025qfal} &
Quantum federated adversarial learning
& Pure
& Centralized \\
\hline

\end{longtable}

Quantum Differential Privacy (QDP)-based QFL systems extend the classical differential privacy technique into quantum computing by making use of quantum information theory and quantum mechanics to enable security. The classical differential privacy adds calibrated randomness to computations or data to make sure that the output does not reveal whether any single individual's data was included. QDP carefully perturbs quantum measurements or quantum states to avoid the risk of disclosure of information about individual quantum states. Techniques like quantum noise, probabilistic quantum measurement, and quantum uncertainty are utilized to create the perturbations. By integration of QDP in QFL, the clients collaboratively train QML models with the guarantee that the individual contributions remain secure from adversaries and also indistinguishable. Recent research reveals that the QDP can ensure stronger privacy guarantees than classical differential privacy. Rofougaran et al. \cite{rofougaran2024federated} combined QFL and QDP for the first time to achieve a framework which is resilient ot both model inversion attacks (due to QDP) and data leakage (due to QFL). The authors also show federated differentially private training as a viable privacy-preserving method. 
Pokharel et al. \cite{pokharel2025differentially} tuned noise variance of QDP through depolarizing channel strength and measurement shots in their QFL system. The authors demonstrated the effectiveness of their framework by considering the relationship between noise parameters and differential privacy budget, and between training accuracy and security.

Blind quantum computing (BQC) allows the clients to delegate quantum computations to computationally stronger quantum servers without disclosing computational tasks or sensitive quantum data. BQC ensures that the server running the computations remains completely unaware of the inputs, intermediate quantum states, and the final output. Quantum techniques like cryptographic obfuscation, unpredictability, and intrinsic uncertainty are employed to achieve blind quantum computing. In QFL systems employing BQC, the client encrypts the (classical) instructions ot manipulating the quantum states along with the input quantum states and communicates them to the server for execution. After the server executes the computations, the client decodes the results. An enhancement of BQC, called verifiable blind quantum computing \cite{drmota2024verifiable}, allows the QFL clients to verify that the server had actually performed the computations as per the provided instructions. Li et al. \cite{li2021quantum} exploited BQC for their QFL system. The authors first proposed a protocol for private single-party training of QML models based on BQC and then extend the same to private multiparty distributed learning with differential privacy. The simulations by the authors showed that the proposed protocol is secure against gradient inversion attack and also robust to experimental imperfections.

\section{Applications}
In this section, we list the applications of QFL across four domains- healthcare, vehicular networks, wireless networks, and network security. A summary of the applications covered is presented in Table \ref{appns}.
\subsection{Healthcare}
QFL for healthcare brings QC's capacity for optimization and compact representations to the privacy-preserving workflow of federated learning. Healthcare is a strong testbed for QFL, given several issues of FL in healthcare like heterogeneous data, limited labels, class imbalance, and the need for robust generalization across hospitals. 

Zhiguo et al. \cite{qu2025daqfl} proposed a Dynamic Aggregation Quantum Federated Learning (DAQFL) framework to improve the performance of QFL models on IoMT-based heterogeneous healthcare data. The framework employs both VQCs and QNNS for local model training. The authors proposed a dynamic aggregation technique where the central server weighs the clients' updates based on the reported accuracy of the client in that round. The higher the accuracy of the local model, the more weight the local model's weights are given in aggregation. The authors conducted experiments on IID, non-IID, and long-tail data distributions, which show the superior performance of the proposed framework in terms of training speed, accuracy, and even with high heterogeneity. 

Balasubramani et al. \cite{balasubramani2025novel} introduced a framework that combined QFL with quantum transfer learning to create a privacy-preserving and robust system for classifying pain levels based on electrocardiogram (ECG) signals. At the EHE computing layer, the raw one-dimensional ECG signals are preprocessed and then transformed into Continuous Wavelet Transform (CWT) images that capture both frequency and time characteristics of cardiac variation. A quantum convolutional hybrid neural network is used to encode the classical features into a 9-qubit PQC for further processing. The authors also used homomorphic encryption to ensure privacy during aggregation of local models. The experimental results show that the proposed QFL model achieved a better performance than the classical model with a tan accuracy of 94.8\%. 

Bhatia et al. \cite{bhatia2023federated} proposed a hybrid quanvolutional neural network-based QFL framework for privacy-preserving training on real-world and non-IID medical image data. The authors use a quanvolutional layer (as a part of QCNN) to encode small subsections of an input image into quantum states using parametrized gates. A learnable quantum circuit is used to process the quantum states, and the resultant outputs' measurements are used for subsequent classical poling and dense layers. The authors also experimented with a combination of a classical CNN with a 16-qubit variational quantum circuit for classification. The experimental results showed that the proposed model achieved a communication speedup of 14 and 40 times over the baseline and also achieved good accuracy. 

In another study, Bhatia et al. \cite{bhatia2024federated} FedQTN- a QFL framework based on Quantum Tensor Networks (QTNs) and differential privacy. The proposed framework utilizes three different hierarchical QTN architectures as local models - multiscale entanglement renormalization ansatz, tree tensor network, and matrix product state. The authors followed a patch-based encoding scheme to encode the input images, where each image is divided into smaller patches that are individually flattened and encoded into quantum states through single-qubit rotations. After encoding, the quantum states are then fed to QTN circuits, which are built using two-qubit unitary gates arranged in a strong or simple entangling block structure. The authors used the FedAvg aggregation algorithm, and the results show that the proposed FedTTN consistently achieved the fastest training convergence. Their experimentation on MRI and CT-scan datasets significantly improved the performance of low-data clients.

\subsection{Vehicular Networks}
QFL for vehicular networks integrates the quantum subroutines with privacy-preserving collaborative training to deal with the heterogeneous and high-velocity roadside data. The vehicular networks setting imposes strict bandwidth and latency limits, intermittent connectivity, and frequent topology changes; hence, integrating light-weight client-side quantum modules with communication-efficient aggregation would be helpful. The importance of security in vehicular networks also motivates quantum-safe keying with quantum protocols.  

Hazarika et al. \cite{hazarika2025quantum} introduced QV-FedCom, a decentralized and heterogeneity-aware QFL framework for the vehicular metaverse that aims to address challenges of data heterogeneity, memory efficiency, and communication cost. The proposed framework has three components - Quantum-inspired PCA (QPCA), Quantum Sequential Training Program (QSTP), and Quantum Vehicle Context Grouping (QVCG). QPCA is applied as a data processing method to compress quantum data and reduce the memory footprint. QVCG manages heterogeneity by employing simulated annealing and hierarchical clustering to group vehicles based on contextual data similarity. QSTP employs reinforcement learning to dynamically switch vehicles between local calibration mode and active streaming mode in order to minimize communication costs. The framework also utilizes quantum trajectory loss (QTL), which is a combination of angular deviation penalty and Huber loss to robustly handle errors in trajectory prediction tasks. The proposed framework consistently outperformed baseline models, achieving a test accuracy of 85\%. The ablation studies also showed the effectiveness of the framework's individual components - QSTP maintained the lowest communication cost, and the integration of QPCA reduced memory consumption from 16GB to 13GB. 

Yamany et al. \cite{yamany2023oqfl} proposed a novel Optimized Quantum-based Federated Learning (OQFL) framework, which is designed to defend against data poisoning attacks in intelligent transportation systems. Quantum-behaved Particle Swarm Optimization (QPSO) is employed by the authors to automatically tune the hyperparameters of the FL process. QPSO optimizes the number of local and global epochs, the learning rate, with an aim of enhancing the model's resilience against adversarial attacks. Analogous to a classical PSO, each particle in QPSO represents a combination of the above-mentioned hyperparameters, and the technique evolves the population to find the combination that yields the highest accuracy. The authors simulated an adversarial environment with four data poisoning attacks while training using QPSO. Tested on the Fashion-MNIST dataset and MNIST dataset, the proposed framework achieved around 98\% accuracy while maintaining a high accuracy of 91\% even when subjected to data poisoning attacks.  

Qu et al. \cite{qu2024qfsm} proposed a novel framework for addressing the challenges of computational efficiency and privacy in speech emotion recognition in 5G Internet of Vehicles (IoV). In the 5G IoV setup, the vehicles collect speech data and send it to edge servers that act as clients in the QFL system. The authors use Quantum Minimal Gated Units (QMGUs) to perform local model training at edge servers. QMGU improves the classical MGU by replacing its non-linear functions with VQCs, which enables faster convergence. The VQCs are implemented using a 4-qubit architecture, which consists of a variational part with RX and CNOT gates and an encoding part with an RX gate. All the edge servers coordinate with a central cloud server for the FL process. The authors show that the proposed QMGU model achieved a superior F1-score compared to QLSTM and QGRU while only using a few resources in terms of qubits and parameters.

\subsection{Wireless Networks}
QFL for wireless networks integrates quantum subroutines into FL to handle tight resource budgets and fast-changing radio environments. The core concerns, like reliability and security, motivate quantum-safe key exchange and robust aggregation against adversarial updates in FL. 

Quy et al. \cite{quy2024federated} explored the application of QFL to Space-Air-Ground Integrated Networks (SAGINs) to enable computationally efficient and privacy-enhanced AI model training for 6G applications. In the proposed framework, a base station serves as a central aggregator with a network of Unmanned Aerial Vehicles (UAVs) being the clients. The proposed quantum machine learning model operates on four qubits, making use of a controlled-rotation (CRX) gate and single-qubit rotation gates (RX, RY, RZ). In the experimentation, the authors compared the performance of their framework with classical FL, showing that the proposed framework has a faster convergence with a convergence of up to 90\% accuracy in 10 epochs, which took classical FL 40 epochs. 

Chaudhary et al. \cite{chaudhary2024quantum} proposed a QFL framework for joint optimization of resource allocation and mode selection in a Simultaneous Transmission and Reflection-Reconfigurable Intelligent Surface (STAR-RIS) assisted Vehicle-Road Cooperation System (VRCS). The authors formulated the joint optimization problem as a Markov Decision Process (MDP), where every V2V pair acts as an agent making decisions. Local observations like queue length and channel gains include the state space for the agent while the action space consists of sub-channel allocation, transmission mode selection, and power control. A novel Quantum Federated Reinforcement Learning (QFRL) algorithm is employed to solve MDP problems where vehicles collaboratively train a global model. A QNN is used as a local model, and each client employs a fidelity-based cost function. The authors demonstrate the superior performance of the proposed QFRL compared to the traditional QFL framework with a 20\% improvement in network sum-rate. QFRL also showed improvements in terms of latency, recording a latency of 2.2 ms compared to 3.2 ms for standard FL.

\subsection{Network Security}
QFL helps in network security by merging with FL with quantum techniques to detect threats across organizations without exposing the edge devices' data and logs to cloud providers. Variational circuits and quantum kernels enrich feature spaces for obfuscated signals, improving sampling efficiency in wireless networks.

Subramanian et al. \cite{subramanian2024hybrid} integrated Quantum-inspired Federated Averaging (QIFA) and Spatio-Temporal Attention Network (STAN) to propose a hybrid federated learning framework for cyber-attack detection. The framework employs hierarchical model aggregation, where the nodes in the network are dynamically grouped into regions based on k-means clustering using a combined score of network conditions and data similarity. After each local node trains its STAN model, the model updates of each region are aggregated by a regional server to create an intermediate regional model. These regional models are then sent to a central server, which uses QIFA for global model aggregation. Notably, the QIFA algorithm introduces periodic quantum-inspired perturbations and a quantum superposition-based adjustment term to avoid local minima and improve convergence. The authors evaluated the proposed framework on the UNSW-NB15 dataset, where it achieved a high accuracy of 98.3\%, which is superior compared to the standard FL models. 

Maouaki et al. \cite{elmaouaki2025qfal} integrated adversarial training into QFL to defend against adversaries or vulnerabilities in a distributed environment. Each client trains a local 6-qubit QNN using amplitude encoding to prepare the initial quantum state. A strongly entangling parametrized circuit with two layers is used to process the encoded quantum states, after which the final measurements are taken for classification. The methodology also involves local adversarial training where a particular fraction of clients generate adversarial examples by using the Projected Gradient Descent (PGD) technique. The clients generating adversarial data update their local models using mini-batches composed of 50\% perturbed data and 50\% clean data. The experimental results showed a clear tradeoff between robustness and accuracy in a 5-client system - an introduction of 20\% adversarial training data improved resistance to attacks by 12.48\%, but at the cost of a 4.83\% drop in clean-data accuracy. The authors noted that larger QFL rounds are more effective in balancing accuracy and robustness.

\section{Case Study 1: Quantum Enhanced Federated Framework for Financial Fraud Detection}
In this section, we present a case for the use of QFL techniques in financial applications in fraud detection, which is derived from the work of Sawaika et al. \cite{sawaika2025privacy}. 

\begin{figure*}[htbp]
    \centering
    \includegraphics[width=0.7\linewidth, height=4in]{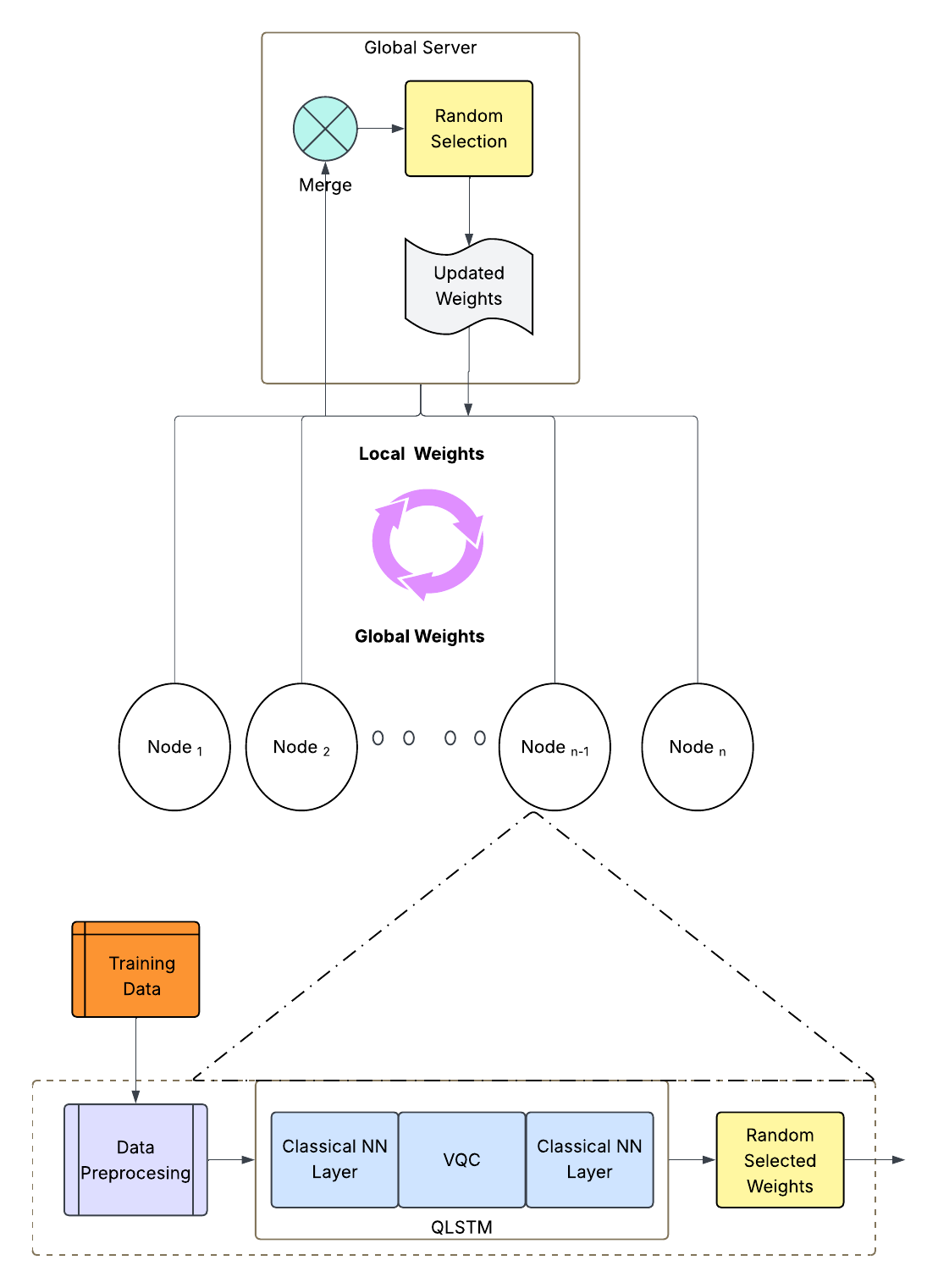}
    \vspace{-0.6cm}
    \caption{\small Training workflow of our framework demonstrating local quantum-enhanced computations, federated aggregation, and secure update exchanges}
    \label{fig:training}
\end{figure*}

\subsection{Motivation}
With an unprecedented increase in the use of digital modes for financial transactions, vulnerabilities to various threats have also increased simultaneously \cite{reurink2019financial}. These include, but are not limited to, insurance fraud, identity theft, phishing, investment fraud, online transaction fraud, etc \cite{sun2023digital}. Some of the recent works using classical ML techniques to solve such problems in financial systems include those by  Hashedi et al. \cite{al2021financial}, Cheng et al. \cite{cheng2024advanced}, and Ali et al. \cite{ali2022financial}. Quantum approaches were also recently studied by Innan et al. \cite{innan2024financial}, Paquet et al. \cite{paquet2022quantumleap}, Innan et al. \cite{innan2024qfnn}, with Innan et al. proposing a federated framework for fraud detection in \cite{innan2024qfnn}. With around a trillion dollars in terms of market impact, McKinsey \cite{gschwendtner2025quantum} predicts around \$633 billion market for quantum technologies in financial use cases. Being one of the critical and sensitive areas of the field, it becomes crucial to investigate state-of-the-art algorithms using quantum federated learning to address such problems for improved performance and accuracy.


\subsection{Problem of Interest}
We designed this QFL model to create a robust framework for fraud detection in online transactional systems. Various design decisions were taken at different stages to create a system that improves performance, via increased accuracy, recall, and reduced false positives. It is also scalable for data and model sizes, and provides for security against various attacks possible in a generalized federated framework. In the solution methodology described in the following section, we will only discuss the modeling strategy for the QFL framework and the underlying ML model used, with a brief discussion on some basic results in the results section.

\subsection{Solution Methodology}

Figure \ref{fig:training} presents an overall training pipeline of the proposed solution with key aspects of the federated framework, the enhanced model, and the privacy-preserving merging strategy. 
The boxes in blue indicate the QLSTM model trained on local data, whereas the yellow ones are part of the merging strategy. The whole system is designed as a pseudo-centralized federated model where the global model is used for aggregation, but doesn't necessarily have the whole information of a consensus model. 

The process starts by preprocessing the dataset. For experiment simulation, we have divided the whole data based on an IID distribution among multiple nodes (parties). But in real-life scenarios, each node will bring its own data.
Each node then preprocesses the data and puts it for training. Once all the local models are trained for one epoch, each node samples the weights and shares only a subset of the weights to the global server. Which then merge the common parameters, sample the merged parameters, and send them back to each node for the next round of training. This sampling is done to minimize the exposure to inference and poisoning attacks. This happens for multiple epochs until an accepted convergence. 
\begin{figure}[htbp]
         \centering
         \includegraphics[width=0.75\linewidth]{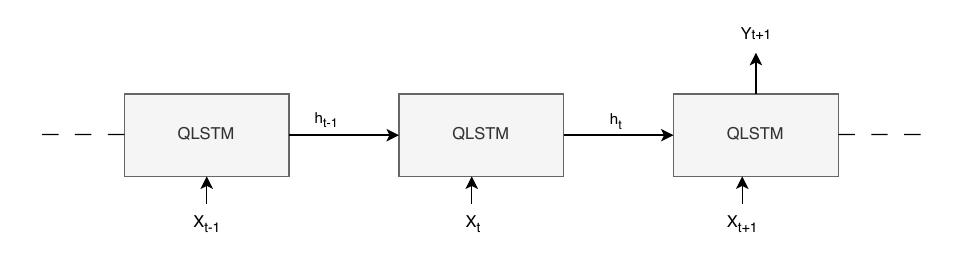}
         \vspace{-0.6cm}
         \caption{QLSTM model demonstrating a 3-sequence architecture with output taken only at the last cell. Each cell processes the input at that time step (t) as $\mathbf{X}_t$ and generates a hidden state $\mathbf{h}_t$, and the output $\mathbf{Y}_{t}$.}
         \label{fig:arc}
\end{figure}

\begin{figure}[htbp]
    \centering
    \includegraphics[width=0.5\linewidth]{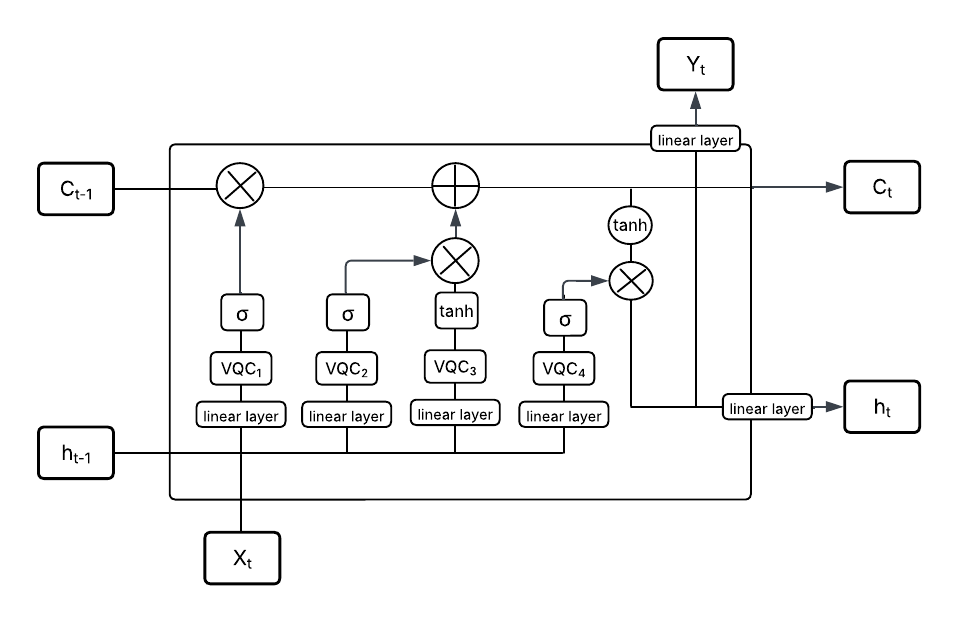}
    \vspace{-0.6cm}
    \caption{\small Architecture of a single QLSTM cell for the model depicted in Figure \ref{fig:arc}.}
    \label{fig:lstm_cell}
\end{figure}

As noted earlier, it consists of a quantum-enhanced LSTM model. In Figure \ref{fig:arc}, we have the structure of a general LSTM model with three cells. Each cell (see Figure \ref{fig:lstm_cell} ) is designed to have four main NN structures, namely forget, input, update, and output. These help in learning the linear structure from the input dataset sequence. Details on the LSTM model can be seen in the original work by Hochreiter et al. \cite{hochreiter1997long}. For each of these functions, we use a VQC to replace a NN.
And as one can see, these VQCs are where our quantum models lie, providing enhancement and guarantees to the problems we set for us in this study.

\begin{figure}[htbp]
    \centering
    \includegraphics[width=\linewidth]{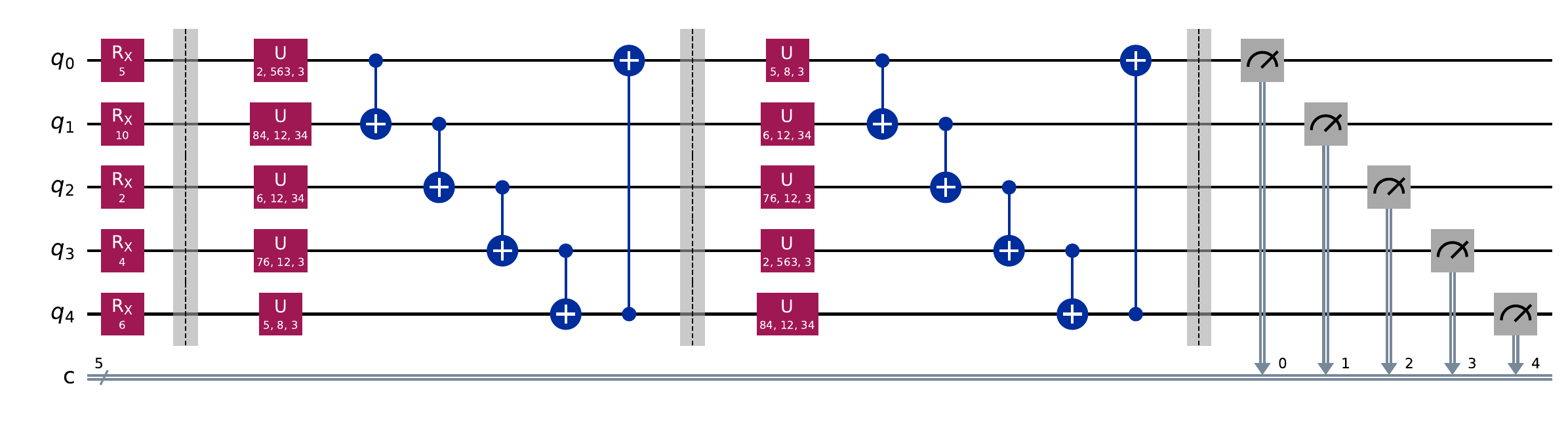}
    \vspace{-0.4cm}
    \caption{ \small A sample five-qubit structure of the designed VQC with two layers. It can be generalized recursively for a larger number of qubits and layers as well.}
    \label{fig:vqc}
\end{figure}

The parameterized structure of the VQC is designed using universal rotation unitaries and {CNOT} gates for full entanglement. Angle encoding using {RX} gates is used as the encoding layer. This design of VQC helps in increasing expressivity of the features in Hilbert's space and to capture complex relations between input features (See Figure \ref{fig:vqc}). The universal {Rot} gate is defined as:
\begin{equation}\label{eq:rot}
R(\phi,\theta,\omega) = RX(\phi)\, RY(\theta)\, RZ(\omega)
\end{equation}

\subsection{Results and Discussion}

This study used a real-life fraud detection bank dataset with 20K rows and 120 encoded numerical features ({Dataset 1}), and a synthetic financial dataset ({Dataset 2}) with 5M rows and five categorical features, available at \cite{Dataset1}, \cite{Dataset2}, respectively. 
The results discussed in this study were obtained on 20K data points from both datasets. Best results are obtained for (number of qubits, depth, sequence length) as (9,10,10) for Dataset 1 and (9, 4, 5) for Dataset 2, respectively.

\begin{figure}[htbp]
    \centering
    \includegraphics[width=0.5\linewidth]{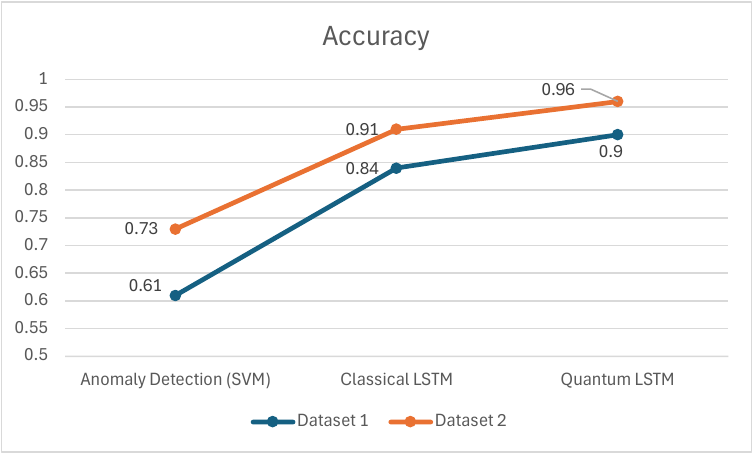}
    \caption{Performance accuracy comparisons of different models under a 5-node federated setup for datasets 1 and 2.}
    \label{fig:Accuracy}
\end{figure}

\begin{figure}[htbp]
    \centering
    \includegraphics[width=0.5\linewidth]{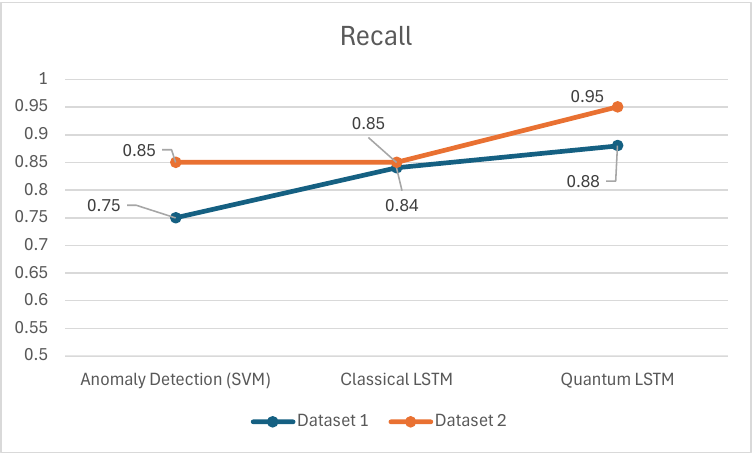}
    \caption{Performance recall comparisons of different models under a 5-node federated setup for datasets 1 and 2.}
    \label{fig:recall}
\end{figure}

\begin{figure}[htbp]
    \centering
    \includegraphics[width=0.5\linewidth]{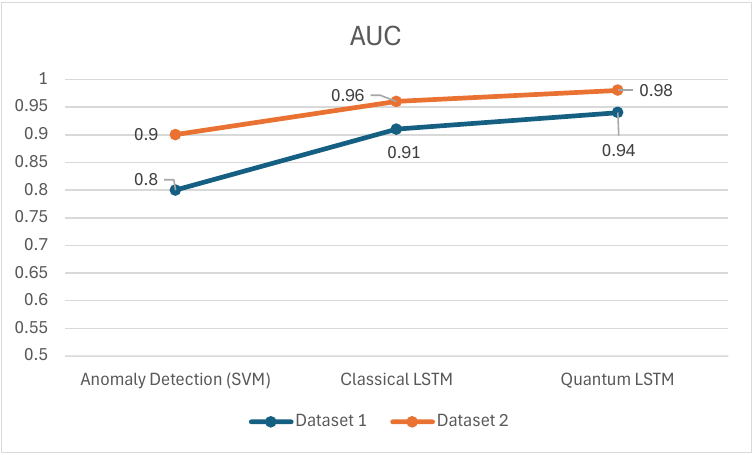}
    \caption{Performance AUC comparisons of different models under a 5-node federated setup for datasets 1 and 2.}
    \label{fig:AUC}
\end{figure}

Figures \ref{fig:Accuracy}, \ref{fig:recall}, \ref{fig:AUC} project the comparison between a classical analogous LSTM model and a traditional non-neural network-based SVM model with the proposed QLSTM model, under a federated setup with five nodes. We observed that One-Class SVM performs poorly, with accuracy scores of only $0.61$ for Dataset 1 and $0.73$ for Dataset 2. More importantly, the QLSTM model outperforms the classical LSTM by approximately 2\% in AUC, 5\% in Accuracy, and 10\% in Recall for Dataset 1; and
by 3\% in AUC, 6\% in Accuracy, and 4\% in Recall for Dataset 2.


\begin{figure}[htbp]
    \centering
    \includegraphics[width=0.9\linewidth]{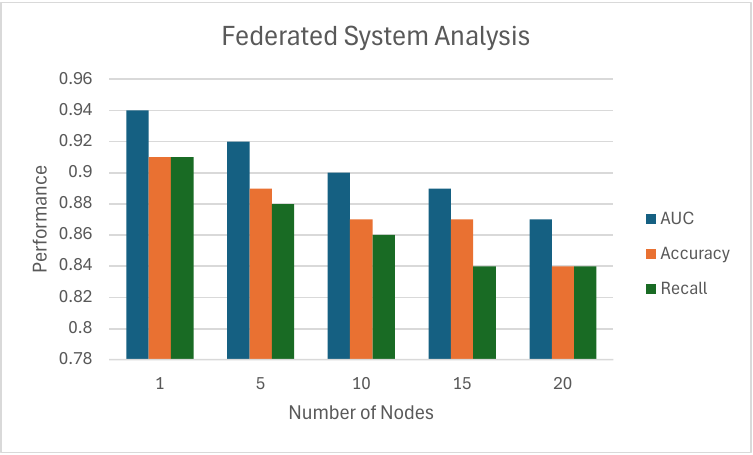}
    \caption{ \small Federation analysis of the QLSTM model for dataset 1}
    \label{fig:n1}
\end{figure}

Figure \ref{fig:n1} plots the performance of the proposed model under federated expansion. We observed a decreasing trend with the increasing number of nodes. This is primarily due to the smaller dataset available per node for training, which limits the local convergence of each model. This may not be the case in real-life scenarios where an additional node brings more data for training in practical use cases. There are mixed results on the impact of node counts on federated performance in the existing literature. But trends shown in figs \ref{fig:n1} are expected for our experiment setup.

\section{Case Study 2 : sat-QFL- Secure Quantum Federated Learning for Low Orbit Satellites}

The popularity of lower Earth orbit (LEO) satellite networks has experienced significant growth over time, with companies such as SpaceX investing substantial resources to facilitate internet accessibility in remote regions. Consequently, researchers have been investigating the potential adaptation of technologies like federated learning and quantum federated learning, which were initially developed for traditional networks, to satellite networks. Satellite networks encounter unique challenges, including fluctuating ground links due to orbital motion, rapidly changing client participation, and stringent latency and energy constraints. These challenges have led to the development of sat-QFL \cite{gurung2025sat}, a hierarchical access-aware quantum federated learning framework. Sat-QFL models satellites and ground stations as a time-varying line-of-sight graph $H(t)$ and assigns two distinct roles: primary satellites that currently possess a direct line of sight to the ground and secondary satellites that must relay data through their neighboring satellites. At any given time $t$, only nodes with a feasible path to a ground station are permitted to participate in the training process. Instead of waiting for all nodes to participate, sat-QFL aggregates the updates that actually arrive. The proposed method, referred to as \emph{sat-QFL}, organizes training sessions around the periods when links are available. It offers three distinct modes for transferring model updates: sequential, simultaneous, and asynchronous. In the sequential mode, secondary satellites transmit weights along a chain until a primary satellite can forward them to the ground station. In the simultaneous mode, multiple secondaries train concurrently and transmit their updates to a primary satellite that performs a local average. In the asynchronous mode, satellites transmit updates whenever a link becomes available, and the receiver aggregates the updates at the conclusion of the link period. These diverse modes contribute to the reduction of idle time caused by strict lockstep rounds and enhance the alignment with the dynamic nature of contact opportunities in orbit.

Security is inherently integrated into the system’s design. Quantum key distribution is employed to establish symmetric keys, and authenticated encryption is utilized to safeguard model exchanges. The study further investigates quantum teleportation as a viable approach for transferring parameter information when prior entanglement is available. These enhancements are intended to provide confidentiality and integrity against quantum-capable eavesdroppers without altering the learning objective or the loss being optimized.
The implementation utilizes Qiskit-based quantum learning workloads and operates on derived constellation traces. Results are presented on the Statlog and EuroSAT datasets, comparing the three scheduling modes and their secure variants, including versions that incorporate teleportation, QKD exclusively, and QKD in conjunction with standard authenticated encryption. The primary trade-off is straightforward. A fully synchronized baseline may appear faster when only round time is considered, but it assumes conditions that Low Earth Orbit (LEO) does not provide. The sat-QFL modes introduce some communication time due to adherence to access windows and the application of security measures, yet they maintain model quality competitiveness and render the overall procedure feasible on orbit. Certain secure runs exhibit variability that warrants further investigation, and the communication-time summaries delineate the anticipated balance between practicality and speed.

In essence, sat-QFL pairs a straightforward role division with window-aware scheduling and built-in security. By grounding participation in $H(t)$ and safeguarding updates over the available links, it transforms quantum federated learning into a process that aligns with the actual connectivity and communication patterns of LEO constellations.


\section{Challenges and Future Research Directions}

\noindent\textbf{Expansion of Machine Learning Tasks:} Most of the research works on QFL focused on classification tasks, thus limiting its potential in other applications \cite{qmi}. Hence, it is crucial to explore beyond classification tasks. Tasks like time-series analysis, optimization, complex decision making, object detection, semantic segmentation, and named entity recognition can greatly benefit from the QFL framework. Expanding QFL applications into these areas could foster innovation in both federated learning and quantum domains, also revealing the advantages of quantum computing. 
Another important challenge in this regard is that the research papers lack a clear justification for using quantum resources - what is the need to use quantum techniques when the same problem is already being efficiently solved by classical methods?
\medskip
\medskip

\noindent\textbf{Quantum techniques for Adversarial attacks:} Although implementing adversarial attacks in QFL setups is more challenging than in FL setups, the recent research shows that it's still possible. Byzantine attacks and gradient inversion attacks are more researched types of adversarial attacks. In Byzantine attacks, malicious clients send corrupted updates for the global model aggregation, while gradient inversion attacks focus on reconstructing clients' data from the gradients shared in the QFL process. Future work can focus on exploiting quantum techniques for building mitigation protocols against adversarial attacks. Existing works are still focusing on classical techniques only to build attack-resistant algorithms.
\medskip
\medskip

\noindent\textbf{Realistic hardware implementation:} Most of the works in the domain of QFL are currently at the level of simulation only. Very few papers explored the hardware implementations of their proposed QFL frameworks. Hence, research must be directed into hardware implementations, considering realistic scenarios like noise. The resilience of the QFL models could be checked by incorporating noise into simulations. Particularly, dephasing or depolarizing noise could be added to local computations. Important inferences on how different types of quantum noise affect global aggregation and local training steps could be inferred, paving the way to real-world deployment of QFL systems.
\medskip
\medskip

\noindent\textbf{Deployment of quantum communication protocols:} Multiple QFL works covered in this chapter utilized quantum key distribution techniques to secure communications. However, the existing quantum hardware is still in a very early stage, making the practical deployment of quantum communication protocols a critical challenge. It requires very costly and sophisticated equipment to create and manipulate entangled particles over long distances. On top of it, building and scaling communication networks in realistic federated learning setups is a great obstacle. 
\medskip
\medskip

\noindent\textbf{Aggregation of different quantum models:} Research works covered in this chapter all aggregate the same type of local models to obtain a global mode. However, in classical FL scenarios, research has been undertaken in terms of the aggregation of different local models. For instance, Scale-FL \cite{ilhan2023scalefl} proposes to aggregate different local models at different scales. Hence, research efforts can be directed to exploring whether different quantum models could learn from one another. If that were to be proved feasible, strategies could be proposed to leverage the unique strength of each quantum model to produce a robust quantum global model.
\medskip
\medskip

\noindent\textbf{Quantum Split Learning as an alternative to QFL:} Split learning is a collaborative learning framework that serves as an alternative to FL, particularly in scenarios with resource-constrained clients. The main idea of split learning is to cut the neural network between clients and server (and then train the neural network in parts) in such a way that there is less computational load on the clients' side. It can be effectively introduced in place of FL to enable Quantum Split Learning. QSL partitions the model so that clients train only the front classical layers or light-weight quantum layers and send the intermediate activations to a server, which does the major part of the quantum layers' training(which is generally computationally heavy). QSL also improves reliability by centralizing the compute-intensive and deeper layers on stable server hardware while allowing heterogeneous clients to train in a flexible fashion. QSL is a better fit when bandwidth is tight, devices are weak, and quantum hardware access is intermittent.

\section{Summary}

This chapter presented a survey of quantum federated learning with a systematic taxonomy covering quantum architecture, data processing method, network topology, and quantum security mechanisms. The design choices across these dimensions depend on the deployment needs of the institutions participating in QFL. The progression from classical FL to QFL shows how the scarce quantum resources should be targeted at high points of leverage like aggregation, secure communication, and feature mapping. Through the representative case study, we validate the end-to-end feasibility of QFL systems. Our findings from the applications of QFL in healthcare, vehicular networks, wireless networks, and network security revealed that the QFL can match or even outperform classical FL models while reducing the time for convergence, and enhancing the communication safety. One of our insights from the recent research in QFL is that the proposed systems are mostly focusing on improving performance compared to classical baselines. However, other dimensions of security and deployment should be researched well to develop practical QFL systems. Another key insight is that split learning in  QFL scenarios would be better suited for resource-limited clients because it can keep all data local, transmit compact activations instead of full gradients, and shift deep computation to resource-heavy servers, thus easing bandwidth constraints. In practical scenarios, the clients in the QFL setup are not expected to have a good amount of quantum computational power. Hence, techniques like shadow tomography should be researched, which delegates the quantum computation to resource-heavy systems. Despite encouraging results, several critical challenges remain in building deployable QFL systems. Apart from realistic hardware deployments, integration of quantum techniques for building adversarial-resistant quantum systems, and the development of standard benchmarks remain a strong direction for future work.

\bibliographystyle{IEEEtran}
\bibliography{main}

\end{document}